\documentclass[twocolumn]{aastex631}  
\usepackage{graphicx}
\usepackage{txfonts}

\usepackage{natbib}
\begin{document}

\title{Evidence of Titanate Clouds in the Day-side Atmosphere of the Ultra-Hot Jupiter WASP-121b}

\author[0000-0001-8018-0264]{Suman Saha}
\affiliation{Instituto de Estudios Astrofísicos, Facultad de Ingeniería Ciencias, Universidad Diego Portales, Av. Ejército Libertador 441, Santiago, Chile}
\affiliation{Centro de Excelencia en Astrofísica y Tecnologías Afines (CATA), Camino El Observatorio 1515, Las Condes, Santiago, Chile}
	
\correspondingauthor{Suman Saha}
\email{suman.saha@mail.udp.cl}

\author[0000-0003-2733-8725]{James S. Jenkins}
\affiliation{Instituto de Estudios Astrofísicos, Facultad de Ingeniería Ciencias, Universidad Diego Portales, Av. Ejército Libertador 441, Santiago, Chile}
\affiliation{Centro de Excelencia en Astrofísica y Tecnologías Afines (CATA), Camino El Observatorio 1515, Las Condes, Santiago, Chile}

\accepted{for publication in The Astrophysical Journal Letters}

\begin{abstract}
The day-side atmospheres of the hottest ultra-hot Jupiters (UHJs) have long been subject to speculation about cloud formation, often without direct observational evidence. Here, we present a detailed analysis of the panchromatic day-side emission spectrum of WASP-121b—one of the hottest known UHJs—covering a broad wavelength range of $\sim$0.6-5.1µm, based on archival JWST observations from NIRISS and NIRSpec/G396H. We report statistically significant detections of several key molecular species, including H$_2$O (13.4 $\sigma$), CO (14.7 $\sigma$), SiO (4.9 $\sigma$), TiO (5.4 $\sigma$), and VO (6.6 $\sigma$), establishing WASP-121b as one of the most thoroughly characterized exoplanetary atmospheres to date. Additionally, we present the robust detection of Titanate (CaTiO$_3$) clouds at 6.7$\sigma$—the first such detection in any exoplanet atmosphere. Our analysis further reveals strong evidence of TiO depletion, likely due to sequestration into refractory condensates such as Titanate clouds. The precisely constrained molecular abundances yield a super-solar C/O ratio of 0.963$\pm$0.024, a sub-solar Si/O ratio of 0.034$\pm$0.024, and a metallicity of 4.7$_{-1.38}^{+1.99}$ $\times$solar. These findings offer a unique window into the atmospheric chemistry of an extreme UHJ, positioning WASP-121b as a key benchmark for next-generation models of atmospheric evolution and dynamics.
\end{abstract}

\section{Introduction}

The hottest ultra-hot Jupiters (UHJs), with dayside temperatures approaching or even exceeding 3000K, serve as exceptional laboratories for studying the most extreme atmospheric evolution in planetary-mass objects \citep[e.g.,][]{2011Natur.469...64M, 2018A&A...617A.110P, 2024ApJS..270...34C, 2024A&A...684A..27D, 2025A&A...700A..45S}. Tidally locked to their host stars and subjected to intense stellar irradiation, their daysides can undergo significant alteration through processes such as photoevaporation \citep[e.g.,][]{2015A&A...576A..42S, 2019A&A...623A..57S, 2022A&A...663A.122C}, photodissociation \citep[e.g.,][]{2018A&A...617A.110P, 2018ApJ...866...27L, 2024A&A...686A..24B}, and contamination by refractory materials \citep[e.g.,][]{2022Natur.604...49L, 2021ApJ...914...12L, 2023ApJ...943..112C}. Moreover, the strong temperature contrast between the day and night sides can drive vigorous atmospheric circulation \citep[e.g.,][]{2019ApJ...886...26T, 2021A&A...653A..73S, 2022AJ....163..155P}, leading to sequestration of condensate species on the cooler night side and, in turn, significantly influencing the dayside atmospheric composition \citep[e.g.,][]{2013A&A...558A..91P, 2022ApJ...934...79K, 2024A&A...685A.139H}.

Given their exceptionally high temperatures and highly inflated radii, UHJs are among the most favorable targets for emission spectroscopy. In particular, the James Webb Space Telescope \citep[JWST,][]{2006SSRv..123..485G}, with its unprecedented spectroscopic capabilities in the near-to-mid-infrared wavelengths, offers a unique opportunity to characterize the atmospheres of these extreme worlds with exceptional precision through both emission and transmission spectroscopy \citep[e.g.,][]{2023Natur.620..292C, 2025AJ....169..341G, 2025NatAs...9..845E, saha2025a, saha2025b}. Day-side emission spectroscopy in the near-to-mid-infrared is especially valuable for detecting molecular absorption and emission features from key species such as H$_2$O, CO, CO$_2$, and SiO. Precise abundance measurements of these constituents enable robust estimation of global atmospheric parameters, including metallicity, the carbon-to-oxygen (C/O) ratio, and refractory-to-volatile ratios (e.g., Si/O). Furthermore, JWST’s coverage extending down to 0.6 µm in the optical range provides critical constraints on the presence and composition of condensate clouds.

The metallicity of the gas giant atmospheres serves as valuable diagnostics for understanding atmospheric mass loss through photoevaporation \citep[e.g.,][]{2018ApJ...865...75N, 2020arXiv200508676M}. However, metallicity can also be affected by disk-driven migration and associated accretion processes, which may enrich the atmospheres of hot-Jupiters and UHJs \citep[e.g.,][]{1998AREPS..26...53B, 2021MNRAS.503.5254L, 2023A&A...675A.176U}. The C/O ratio is likewise widely regarded as a key tracer of giant planet formation and migration pathways \citep{2011ApJ...743L..16O, 2018A&A...613A..14E, 2019A&A...632A..63C}. Super-solar C/O ratios inferred in several UHJs have been interpreted as signatures of formation beyond the snow line, followed by inward migration \citep{2011Natur.469...64M, 2024AJ....168...14W, 2024AJ....168..293S}. However, the relatively small number of precise C/O measurements currently available limits our ability to robustly constrain such formation scenarios. The C/O ratio has also been proposed as a diagnostic for distinguishing between formation via core accretion and gravitational instability \citep{2014ApJ...794L..12M, 2024ApJ...969L..21B}. In addition, oxygen depletion due to sequestration into refractory condensates has been suggested as a mechanism that could artificially elevate the observed C/O ratios in gas giant atmospheres \citep{2023MNRAS.520.4683F}. Beyond the C/O ratio, refractory-to-volatile abundance ratios—such as Si/O—have recently emerged as valuable tracers of planetary formation and evolutionary history \citep[e.g.,][]{2021ApJ...914...12L, 2022A&A...665L...5K, 2023ApJ...943..112C}.

Since its discovery in 2016, WASP-121b \citep{2016MNRAS.458.4025D} has become one of the most extensively studied UHJs. As one of the hottest (T$_{eq}$ $\sim$ 2358K) and most inflated (R$_{p}$ $\sim$ 20.9 R$_\oplus$) UHJs known, it has been classified as an extreme ultra-hot Jupiter (EUHJ) by \citet{saha2025b}, a category that currently includes only six known exoplanets. Low-resolution transmission spectroscopy with HST and Gemini-S/GMOS has suggested possible atmospheric variability \citep{2021MNRAS.503.4787W, 2024ApJS..270...34C}. In addition, numerous transmission and emission spectroscopic studies using both ground- and space-based facilities have been conducted to characterize its extreme atmosphere.

High resolution transmission spectroscopy using HARPS \citep{2020ApJ...897L...5B, 2020MNRAS.494..363C, 2020A&A...641A.123H}, VLT/UVES \citep{2020MNRAS.493.2215G, 2021MNRAS.506.3853M}, and VLT/ESPRESSO \cite{2021A&A...645A..24B, 2022A&A...666L..10A, 2023MNRAS.519.1030M, 2024A&A...685A.139H}—including observations in the 4UT mode \citep{2025A&A...694A.284P}—has revealed several atomic and ionic features in the atmosphere of WASP-121b, such as H I, Li I, Na I, K I, Mg I, Ca I, Ti I, etc. Low-resolution transmission spectroscopy with HST/WFC3 and HST/STIS \cite{2016ApJ...822L...4E, 2018AJ....156..283E} has detected H$_2$O and VO, whereas low-resolution emission spectroscopy with HST/WFC3 \citep{2020MNRAS.496.1638M} has reported the presence of H$_2$O in the day-side atmosphere. Phase resolved transmission spectroscopy with Gemini-S/IGRINS \cite{2024PASP..136h4403W} detected both CO and H$_2$O features. High-resolution emission spectroscopy using Gemini-S/IGRINS \citep{2024AJ....168..293S}, as well as VLT/ESPRESSO and CRIRES+ \citep{2025AJ....169...10P}, confirmed the presence of CO and H$_2$O, and suggested a super-solar C/O ratio. More recently, transmission and emission spectroscopy with JWST/G395H has detected H$_2$O, CO, and SiO in the atmosphere, along with additional evidence of a super-solar C/O ratio \citep{2025AJ....169..341G, 2025NatAs...9..845E}.

In this work, we present an analysis of the panchromatic emission spectrum of WASP-121b using the archival JWST observations from NIRISS and NIRSpec/G395H. The broad wavelength coverage and high signal-to-noise ratio (S/N) of the combined dataset enable us to precisely constrain the dayside abundances of key atmospheric constituents, including the cloud properties. In Section \ref{sec:sec2}, we detail our methodology, including data reduction, light curve analysis, and atmospheric retrievals. In Section \ref{sec:sec3}, we present our key findings and discuss their broader implications in detail.

\section{Methodology}\label{sec:sec2}

\subsection{Data reduction and analysis}\label{sec:sec2.1}

WASP-121b was observed by both the NIRISS \citep{2023PASP..135i8001D} and NIRSpec/G395H \citep{2022A&A...661A..80J} instruments on board JWST as part of programs $\#$1201 (PI: David Lafreniere) and  $\#$1729 (PI: Thomas Evans-Soma). Both observations covered full phase curves of WASP-121b, including two secondary eclipse events in each case. Together, these two instruments provide complementary wavelength coverage, delivering medium-resolution spectra spanning approximately 0.6-5.1µm. We accessed the publicly available archival data via the Barbara A. Mikulski Archive for Space Telescopes (MAST)\footnote{\url{https://mast.stsci.edu/}} \dataset[(doi: 10.17909/ns2e-df07)]{\doi{10.17909/ns2e-df07}}. The NIRISS data were divided into 19 segments, of which segments 1-5 and 15-19, covering the two secondary eclipses, were used in our analysis. Similarly, the NIRSpec data were divided into 13 segments, and we used segments 1–3 and 11–13 from both the NRS1 and NRS2 detectors.

The raw data (\texttt{*uncal.fits}) from the JWST Science Calibration Pipeline \citep{2022zndo...7071140B} for both observations were processed using two independent community-developed pipelines: \texttt{Eureka!} \citep{Bell2022} and \texttt{exoTEDRF} \citep[formerly \texttt{supreme-SPOON},][]{2024JOSS....9.6898R}. Both pipelines are widely adopted and have been rigorously tested for accuracy, particularly \texttt{Eureka!}, as demonstrated in several recent studies (e.g., \citealt{2023Natur.614..664A, 2023NatAs...7.1317L, 2024Natur.626..979P, 2023ApJ...959L...9M} for \texttt{Eureka!}; and \citealt{2024ApJ...962L..20R, 2023Natur.614..670F} for \texttt{exoTEDRF}). In addition to the science frames, these reduction pipelines require calibration files, which we retrieved from the JWST Calibration Reference Data System (CRDS)\footnote{\url{https://jwst-crds.stsci.edu/}}. Employing two independent pipelines allows us to cross-validate our reductions and analyses by comparing results, an important step for ensuring robustness in high-precision studies such as this.

The \texttt{Eureka!} pipeline offers a six-stage procedure for reducing and analyzing JWST time-series data across various instruments and observing modes, with each stage comprising multiple built-in functions and subprocesses \citep{Bell2022}. For this study, we employed the first three stages—covering calibration, detector-level correction, and optimal spectral extraction. The NRS1 and NRS2 datasets from NIRSpec were processed independently. Our reductions largely followed the default configurations specified in the instrument-specific “\texttt{*.ecf}” files for both NIRISS and NIRSpec/G395 spectral reductions, with only minor modifications to improve performance, as detailed in \cite{saha2025b}. Similarly, \texttt{exoTEDRF} offers a streamlined framework for JWST raw data reduction across multiple instruments, structured into four processing stages. We utilized the first three stages, corresponding to calibration, corrections, and spectral extraction. In line with the \texttt{Eureka!} workflow, the NIRSpec NRS1 and NRS2 datasets from NIRSpec were treated separately. We primarily adhered to the default configuration settings optimized for NIRISS and NIRSpec/G395H, applying only minor adjustments as outlined in \cite{saha2025b}.

\begin{figure}
\centering
\includegraphics[width=\columnwidth]{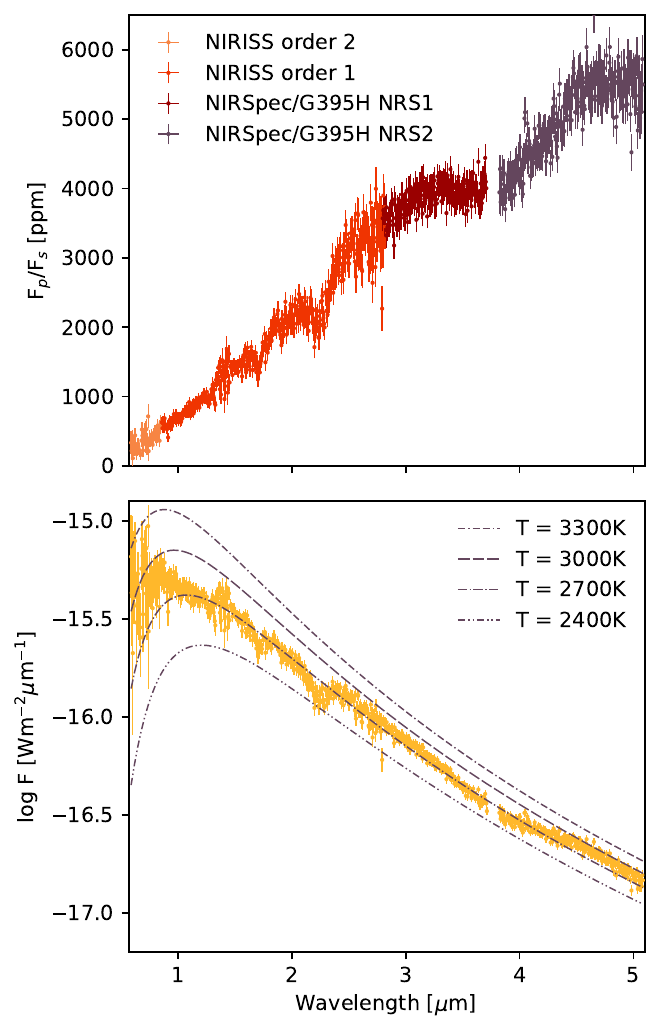}
\caption{(Upper) The observed panchromatic planet-to-star flux ratio of WASP-121b, spanning a wavelength range of $\sim$0.6- 5.1µm, extracted using the \texttt{Eureka!} pipeline from JWST NIRISS and NIRSpec/G395H observations, followed by spectral light curve fitting with \texttt{ExoELF}. Contributions from individual datasets across the wavelength range are shown in different colors. (Lower) The panchromatic emission spectral energy distribution (SED) of WASP-121b, derived from the planet-to-star flux ratio and a \texttt{PHEONIX} stellar model. Overplotted blackbody curves corresponding to different brightness temperatures illustrate spectral features in the planet's spectrum. The significant deviation of the observed spectra from blackbody expectations at shorter wavelengths indicates the increasing dominance of starlight reflected by the planet in this regime. \label{fig:fig1}}
\end{figure}

\subsubsection{Lightcurve modeling}\label{sec:sec2.2}

The extracted spectroscopic time-series data (i.e., 2D lightcurves) from both the \texttt{Eureka!} and \texttt{exoTEDRF} reductions were binned along the wavelength axis to produce the spectroscopic lightcurves. We tested multiple binning schemes and, given the high signal-to-noise ratio (S/N) of our observations, adopted a relatively high-resolution approach using uniformly spaced 0.005 µm bins. This corresponds to a spectral resolution of $\sim$250-2000 across the wavelength range. The final binning yielded a total of 879 spectroscopic lightcurves: 54 from NIRISS Order 2, 388 from NIRISS Order 1, 184 from NIRSpec NRS1, and 253 from NIRSpec NRS2.

The lightcurves were modeled using our in-house pipeline, \texttt{ExoELF} \citep[ExoplanEts Lightcurves Fitter,][]{saha2025a, saha2025b, 2025MNRAS.539..928S}, which integrates several widely adopted packages. These include \texttt{batman} \citep{2015PASP..127.1161K} for transit and eclipse modeling, \texttt{emcee} \citep{2013PASP..125..306F} and \texttt{DYNESTY} \citep{2020MNRAS.493.3132S} for MCMC and nested sampling, respectively, and \texttt{celerite} \citep{celerite1, celerite2} and \texttt{george} \citep{2015ITPAM..38..252A} for Gaussian process (GP) regression. \texttt{ExoELF} is optimized for fitting large volumes of light curves, such as those generated from JWST transmission and emission spectroscopy, and supports flexible detrending strategies tailored to high-precision datasets.

We adopted a two-stage procedure to extract the emission spectrum from the light curves. In the first stage, we simultaneously modeled the white light curves from NIRISS Order 1, and NIRSpec NRS1 and NRS2, treating the orbital parameters and mid-transit times as free parameters (see Table \ref{tab:lc_w} and Figure \ref{fig:lc_w}). NIRISS Order 2 was excluded from this fit due to its significantly lower S/N. In the second stage, the orbital parameters and mid-transit times obtained from the white light curve analysis were fixed, and we modeled the spectroscopic light curves by fitting only for the eclipse depth and detrending parameters (see Figure \ref{fig:fig1}). We tested several detrending schemes extensively and ultimately adopted a two-fourth-order polynomial approach. This configuration allowed us to effectively detrend both eclipses in each light curve with minimal parameter overhead, while preserving a likelihood function comparable to that of more complex detrending schemes. All joint light curve fittings were performed using nested sampling via \texttt{DYNESTY}.

The emission spectra obtained from both independent data reduction pipelines—\texttt{Eureka!} and \texttt{exoTEDRF}—show a mean absolute difference of only $\sim$0.5$\sigma$ (see Figure \ref{fig:spec_diff}), demonstrating excellent statistical agreement and affirming the robustness of our extracted spectra. However, consistent with our previous findings \citep{saha2025a, saha2025b}, we observed that the \texttt{exoTEDRF} spectrum exhibits notably smaller estimated uncertainties, despite lacking a corresponding reduction in variance relative to the \texttt{Eureka!} spectrum. Consequently, we adopt the \texttt{Eureka!} spectrum as the default dataset for our atmospheric retrieval analyses, while still performing all analyses on both datasets to cross-validate the robustness of our reported results. Furthermore, when compared with several previous studies based on NIRSpec/G395H, HST, TESS, TRAPPIST, and SMARTS observations, our panchromatic emission spectra are broadly consistent \citep{2025NatAs...9..845E, 2022NatAs...6..471M, 2020A&A...637A..36B, 2020MNRAS.496.1638M, 2016MNRAS.458.4025D}.

\begin{figure*}
\centering
\includegraphics[width=2\columnwidth]{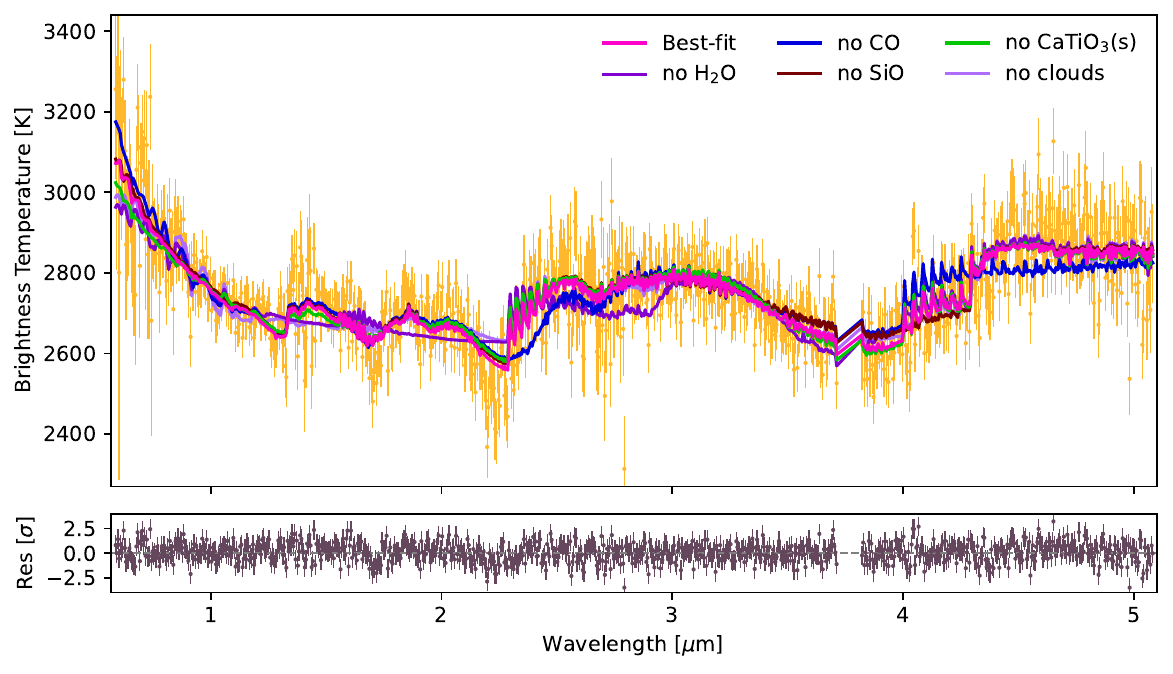}
\caption{The observed panchromatic emission spectrum of WASP-121b (from the \texttt{Eureka!} reduction), shown in terms of brightness temperature, overplotted with the best fit free-chemistry model from the retrieval analyses, along with the corresponding residuals. Additional retrieved models, each generated by excluding specific species, are also overplotted; these were used to assess the detection significance of those species. \label{fig:fig2}}
\end{figure*}

\begin{table*}
    \centering
    \caption{Retrieved molecular abundances (in log MMRs), CaTiO$_3$(s) cloud-base abundance and sedimentation efficiency, along with the corresponding model log-evidences from the atmospheric retrieval analyses.}
    \label{tab:tab1}
    \[
    \begin{array}{lccccccccc}
        \hline
        \hline
        \text{Case} & [\mathrm{H}_2\mathrm{O}] & [\mathrm{CO}] & [\mathrm{SiO}] & [\mathrm{HCN}] & [\mathrm{TiO}] & [\mathrm{VO}] & [\mathrm{CaTiO_{3}(s)}]^{\dagger} & \mathrm{F_{sed}} & \ln Z \\
        \hline
        \text{Eureka! (free)} &
        -4.13_{-0.09}^{+0.09} &
        -1.37_{-0.14}^{+0.14} &
        -2.62_{-0.30}^{+0.26} &
        -5.63_{-4.13}^{+0.29} &
        -11.40_{-1.41}^{+1.38} &
        -9.46_{-2.18}^{+0.86} &
        -1.80_{-0.15}^{+0.16} &
        7.02_{-1.21}^{+1.26} &
        33571.53 \\
        \text{Eureka! (free/no CaTiO}_3\text{(s))} &
        -4.09_{-0.06}^{+0.06} &
        -1.24_{-0.10}^{+0.10} &
        -2.30_{-0.16}^{+0.15} &
        -5.00_{-0.13}^{+0.11} &
        -8.99_{-1.22}^{+1.10} &
        -8.10_{-0.41}^{+0.34} &
        ... & ... &
        33549.15 \\
        \text{Eureka! (free/no clouds)} &
        -4.26_{-9.11}^{+0.11} &
        -1.40_{-0.17}^{+0.16} &
        -2.52_{-0.30}^{+0.28} &
        -5.02_{-0.16}^{+0.14} &
        -6.31_{-0.14}^{+0.14} &
        -7.01_{-0.15}^{+0.14} &
        ... & ... &
        33519.78 \\
        \text{Eureka! (equilibrium)} &
        -3.45_{-0.06}^{+0.08} &
        -1.79_{-0.11}^{+0.15} &
        -3.05_{-0.12}^{+0.16} &
        -4.07_{-0.12}^{+0.16} &
        -5.96_{-0.14}^{+0.19} &
        -8.15_{-0.16}^{+0.22} &
        -2.04_{-0.19}^{+0.14} &
        6.70_{-2.10}^{+1.79} &
        33561.49 \\
        \text{Eureka! (equilibrium/no CaTiO}_3\text{(s))} &
        -3.31_{-0.04}^{+0.04} &
        -1.04_{-0.08}^{+0.08} &
        -2.29_{-0.09}^{+0.09} &
        -3.20_{-0.09}^{+0.09} &
        -5.44_{-0.12}^{+0.12} &
        -7.66_{-0.13}^{+0.13} &
        ... & ... &
        33548.17 \\
        \text{Eureka! (equilibrium/no clouds)} &
        -3.77_{-0.04}^{+0.04} &
        -1.93_{-0.07}^{+0.07} &
        -3.51_{-0.09}^{+0.09} &
        -4.10_{-0.08}^{+0.08} &
        -7.02_{-0.10}^{+0.10} &
        -9.24_{-0.10}^{+0.11} &
        ... & ... &
        33490.53 \\
        \text{exoTEDRF (free)} &
        -4.27_{-0.07}^{+0.08} &
        -1.56_{-0.16}^{+0.18} &
        -3.21_{-0.19}^{+0.20} &
        -5.57_{-0.14}^{+0.13} &
        -11.42_{-1.44}^{+1.46} &
        -7.99_{-0.11}^{+0.11} &
        -0.22_{-0.71}^{+0.15} &
        9.16_{-1.22}^{+0.50} &
        33520.47 \\
        \text{exoTEDRF (equilibrium)} &
        -3.61_{-0.02}^{+0.02} &
        -2.23_{-0.03}^{+0.04} &
        -3.37_{-0.03}^{+0.04} &
        -4.79_{-0.05}^{+0.05} &
        -6.40_{-0.03}^{+0.05} &
        -8.69_{-0.04}^{+0.06} &
        -0.58_{-0.48}^{+0.35} &
        4.11_{-1.27}^{+1.84} &
        33495.13 \\
        \hline
        \hline
        \text{${}^\dagger$ Abundance at the cloud-base.}
    \end{array}
    \]
\end{table*}

\subsection{Atmospheric retrieval analyses}\label{sec:sec2.3}

Owing to the excellent S/N in the extracted emission spectra, we performed a detailed atmospheric characterization through retrieval analyses. For this purpose, we adopted \texttt{petitRADTRANS} \citep[\texttt{pRT},][]{2019A&A...627A..67M}, a widely used and well-validated radiative transfer code within the exoplanet community \citep[e.g.,][]{2024arXiv241008116K, 2024ApJ...973L..41I, 2024arXiv241203675L, 2024MNRAS.527.7079P, 2024A&A...690A..63B}. Although pRT employs a one-dimensional radiative transfer formalism, it remains sufficiently robust for the reliable interpretation of exoplanet spectra observed with current instruments such as JWST, offering nearly all functionalities required for such complex analyses.

Our retrieval framework is based on the \texttt{pRT} retrieval routines \citep{2024JOSS....9.5875N}, which incorporates \texttt{PyMultinest} \citep{2009MNRAS.398.1601F, 2014A&A...564A.125B} for nested sampling of the posterior space. Given the high spectral resolution of our extracted emission spectra, we used the line-by-line opacity mode, setting the model resolution to 6000, which is optimal for the spectral resolution of our data. We adopted the pressure-temperature (PT) profile parametrization from \citet{2010A&A...520A..27G}, which provides sufficient flexibility without unnecessary computational complexity. This parametrization is governed by four variables: the equilibrium (irradiation) temperature (T${\mathrm{eq}}$), the internal temperature (T${\mathrm{int}}$), the visible-to-infrared opacity ratio ($\gamma$), and the mean infrared opacity ($\kappa_{\mathrm{IR}}$).

We employed both free and equilibrium chemistry models to describe the abundances of chemical species in our retrieval analyses. In the free chemistry approach, the abundances (expressed as log-MMRs) of individual species were treated as vertically uniform free parameters, typically ranging between -14 and -0.5 (see Table \ref{tab:retrieval_priors}). To improve computational efficiency, the upper bound was lowered for most species that did not show evidence of requiring a higher range in preliminary fits. For the equilibrium chemistry models, we used precomputed abundance tables provided by \texttt{pRT}, assuming chemical equilibrium at each pressure level.

We included the opacities of all key molecules expected to contribute significantly to the emission spectra across the observed wavelengths, including H$_2$O \citep{2018MNRAS.480.2597P}, CO \citep{2010JQSRT.111.2139R}, CO$_2$ \citep{2010JQSRT.111.2139R}, SiO \citep{2013MNRAS.434.1469B}, CH$_4$ \citep{2020ApJS..247...55H}, HCN \citep{2006MNRAS.367..400H}, C$_2$H$_2$ \citep{2013JQSRT.130....4R}, PH$_3$ \citep{2015MNRAS.446.2337S}, H$_2$S \citep{2013JQSRT.130....4R}, TiO \citep{2019A&A...627A..67M}, and VO \citep{2019A&A...627A..67M}. Rayleigh scattering by H$_2$ \citep{1962ApJ...136..690D} and He \citep{1965PPS....85..227C} was included, along with collision-induced absorption (CIA) from H$_2$-H$_2$ \citep{2001JQSRT..68..235B, 2002A&A...390..779B} and H$_2$-He \citep{1988ApJ...326..509B, 1989ApJ...336..495B}. Additionally, bound-free and free-free absorption H$^-$ \citep{2008oasp.book.....G} were incorporated into our equilibrium chemistry model, as these processes can be significant at the high temperatures typical of UHJ atmospheres.

We also included several condensate species expected to form clouds at these temperatures, following the formalism by \citet{2001ApJ...556..872A}, which include Fe(s) \citep{2018MNRAS.475...94K, 1994ApJ...421..615P, 1997A&A...327..743H, 2001ApJ...556..872A}, MgSiO$_3$(s) \citep{1994A&A...292..641J, 2001ApJ...556..872A}, Al$_2$O$_3$(s) \citep{1995Icar..114..203K, Stull1947}, SiO$_2$(s) \citep{2018MNRAS.475...94K, 1997A&A...327..743H, 1995Icar..114..203K, Stull1947}, TiO$_2$(s) \citep{2018MNRAS.475...94K, 2011A&A...526A..68Z, 2003ApJS..149..437P, 2016arXiv160704866S, 2019RJPCA..93.1024S}, and CaTiO$_3$(s) \citep{2018MNRAS.475...94K, 2003ApJS..149..437P, 1998JPCM...10.3669U, 2019RJPCA..93.1024S}. The abundances of these species at the cloud base were treated as free parameters in both free and equilibrium chemistry models. In addition, we treated the sedimentation efficiency  (f$_{\mathrm{sed}}$), log-normal width of particle size distribution ($\sigma$), the vertical eddy diffusion coefficient (k$_{\mathrm{zz}}$), and the cloud fraction (f$_{\mathrm{c}}$) as additional free parameters, which collectively constrain the cloud profiles \citep{2001ApJ...556..872A}. We also included day-side scattering of starlight, using a day-side averaged irradiance.

\begin{figure}
\centering
\includegraphics[width=1\columnwidth]{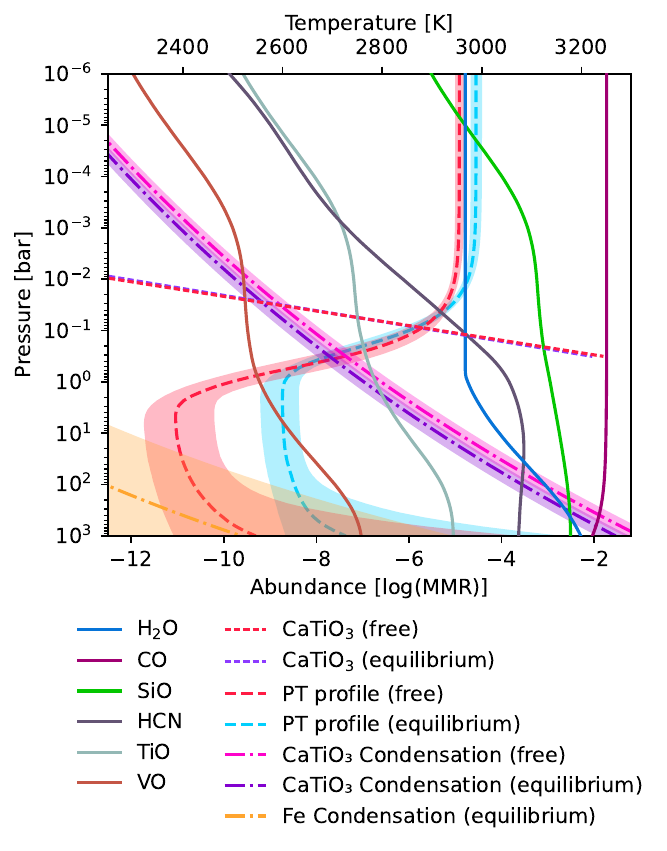}
\caption{Pressure-dependent median abundances (in log-MMRs) from the best-fit equilibrium chemistry retrieval of the observed emission spectrum of WASP-121b (from \texttt{Eureka!}), along with the median abundance profiles of the condensate species CaTiO\textsubscript{3}}. Overplotted are the pressure-temperature (PT) profiles from the best-fit retrievals using both free and equilibrium chemistry models, highlighting their close agreement. The condensation curves for CaTiO\textsubscript{3} from both retrieval approaches are also shown, along with the Fe condensation curve from the equilibrium chemistry retrieval. \label{fig:fig3}
\end{figure}

\section{Results and Discussion}\label{sec:sec3}

The high S/N panchromatic emission spectra of WASP-121b, spanning over a broad wavelength range of $\sim$0.6-5.1µm, enabled a highly constraining and precise atmospheric characterization through our retrieval analyses. Both the free and equilibrium chemistry retrievals achieved excellent fits to the observed data.

From the free chemistry retrieval of the \texttt{Eureka!} spectra, we obtained precise constraints on the abundances of H$_2$O, CO, SiO, and TiO, and marginal constraints on HCN and VO (see Figure \ref{fig:fig2}, Table \ref{tab:tab1}, and Figure \ref{fig:cor_eu_f}). The retrievals also constrained the abundance of CaTiO$_3$(s) condensates, indicating its presence in the atmosphere. To assess the statistical significance for the detection of individual species, we performed retrievals excluding each species from the model one at a time (see Figure \ref{fig:fig2}, \ref{fig:fig2a}). From these comparisons, we report statistically significant detection of H$_2$O (13.37 $\sigma$), CO (14.72 $\sigma$), SiO (4.86 $\sigma$), TiO (5.37 $\sigma$), and VO (6.61 $\sigma$), along with a tentative detection of HCN (0.89 $\sigma$).

These results make WASP-121b the first exoplanet with statistically significant detections and well-constrained abundances for three distinct refractory species—SiO, TiO, and VO. The retrievals also show a strong detection of clouds in the day-side atmosphere (10.17 $\sigma$), including a significant detection of Titanate (CaTiO$_3$(s), also known as Perovskite) clouds at 6.69 $\sigma$ (see Figure \ref{fig:fig3}). This marks the first reported detection of Titanate clouds in an exoplanetary atmosphere, consistent with expectations for the hottest planetary environments. WASP-121b thus becomes only the second ultra-hot Jupiter, after WASP-19b \citep{saha2025a}, with a statistically significant detection of clouds.

To verify these findings, we also performed free chemistry retrievals on the \texttt{exoTEDRF} spectrum. The results showed consistent abundances, while also providing more precise constraints for both HCN and VO—likely due to the lower uncertainty levels in the \texttt{exoTEDRF} spectrum (see Table \ref{tab:tab1} and Figure \ref{fig:cor_et_f}). This analysis also confirmed a significant detection of clouds at 12.46 $\sigma$, including a 9.07 $\sigma$ detection of CaTiO$_3$(s) condensate clouds.

The equilibrium chemistry retrievals of the \texttt{Eureka!} spectra—using the same set of molecular species detected in the free chemistry retrieval—resulted in well-constrained abundances for all included species (see Table \ref{tab:tab1} and Figure \ref{fig:cor_eu_eq}). Retrievals using exclusion models again demonstrated statistically significant detections for H$_2$O (12.14$\sigma$), CO (16.12$\sigma$), and SiO (5.19$\sigma$), along with marginal detections for TiO (3.09$\sigma$) and HCN (1.51$\sigma$).

Interestingly, the TiO abundance inferred from the equilibrium chemistry retrieval is significantly higher than that derived from the free chemistry retrieval. This discrepancy suggests that the low TiO abundance observed in the free chemistry model is inconsistent with chemical equilibrium, providing strong evidence for the sequestration of TiO via cloud condensation. This represents the first statistically supported case in exoplanetary science for the depletion of a refractory species—specifically TiO—through condensation, as the best-fit free chemistry model is favored over the equilibrium chemistry model at 4.48$\sigma$. Notably, the elevated TiO abundance in the equilibrium chemistry retrieval leads to the statistical non-detection of VO, in contrast with the free chemistry results.

Nonetheless, consistent with the free chemistry analysis, the equilibrium chemistry retrieval shows strong evidence for clouds in the dayside atmosphere (11.91$\sigma$), including a significant detection of Titanate (CaTiO$_3$(s)) clouds at 5.16$\sigma$ (see Figure \ref{fig:fig3}). It also suggests the possible presence of iron (Fe(s)) clouds, as indicated by a broadly constrained Fe(s) abundance (see Figures \ref{fig:fig3} and \ref{fig:cor_eu_eq}).

To compare our retrieved properties with those reported by \citet{2025NatAs...9..845E}, who also detected H$_2$O, CO, and SiO using NIRSpec/G395H observations, we converted the median abundances from our free chemistry retrieval of the Eureka! spectra into VMRs (H$_2$O$_{\rm VRM}$ = -5.01, CO$_{\rm VRM}$ = -2.44, SiO$_{\rm VRM}$ = -3.88). Their reported abundances are broadly consistent with ours within 1-2$\sigma$, although our estimates are systematically slightly lower. While their retrieved PT profile generally matches ours in temperature range and general structure, our retrievals indicate a temperature inversion occurring deeper in the atmosphere—likely a better-constrained result owing to our broader spectral coverage, particularly at shorter wavelengths, and the inclusion of realistic cloud models.

We also note that both TiO and VO were not robustly detected in previous studies using high-resolution ground-based instruments \citep[e.g.,][]{2020A&A...636A.117M, 2025A&A...694A.284P}, and TiO has been suggested to be cold-trapped on the night-side atmosphere \citep{2020A&A...641A.123H, 2024A&A...685A.139H}. Our study presents the first robust detection of both molecules in this atmosphere, albeit at depleted levels. Additionally, \citet{2022NatAs...6..471M} previously discussed the possibility of CaTiO$_3$ condensation on the night-side of WASP-121b, whereas our retrievals indicate its presence on the dayside.

We also investigated potential H$_2$O thermal dissociation in the upper atmospheres by performing additional free chemistry retrievals where H$_2$O abundance scales as P$^{\tau}$ above a reference pressure P$_0$, following \citet{2025NatAs...9..845E}. We found no significant evidence for dissociation: $\tau$ remained highly unconstrained, P$_0$ was driven below $10^{-3}$ bar, and the log-evidence decreased compared to the standard model. Moreover, the retrieved H$_2$O abundances are broadly consistent between our free and equilibrium chemistry models in the upper atmosphere, which reinforces this conclusion.

Using the precisely constrained molecular abundances from the free chemistry retrieval of the \texttt{Eureka!} spectra, we derived a super-solar C/O ratio of 0.963$\pm$0.024. This robust and exceptionally precise estimate—enabled by the well-constrained abundances of several key molecular species—highlights JWST’s transformative capability to probe UHJ atmospheres via emission spectroscopy. Additionally, molecular abundances obtained from the equilibrium chemistry retrieval of the \texttt{Eureka!} spectra yielded a consistent C/O ratio of 0.941$\pm$0.02.

The super-solar C/O ratio estimated for WASP-121b is consistent with those observed for several other UHJs \citep[e.g.,][]{2011Natur.469...64M, 2024AJ....168...14W, saha2025a, saha2025b}, suggesting a possible signature of shared evolutionary pathways or high-temperature chemistry unique to this class of planets. A C/O ratio above solar is often interpreted as evidence that the planet formed beyond the ice line and subsequently migrated inward to its current close-in orbit \citep{2017AJ....153...83B, 2021A&A...654A..71S}. However, material accretion can obscure distinctions between formation via core accretion and gravitational instability \citep{2014ApJ...794L..12M, 2024ApJ...969L..21B}. Furthermore, as proposed by \cite{2023MNRAS.520.4683F}, an elevated C/O ratio may also result from oxygen depletion due to sequestration into refractory condensates. In the case of WASP-121b, the observed depletion of TiO, as discussed earlier, points to possible oxygen removal through the formation and rainout of heavy condensates such as Titanate clouds. Such distinctive observational signatures provide critical insights into the chemical and dynamical processes shaping UHJ atmospheres.

\begin{figure}
\centering
\includegraphics[width=\columnwidth]{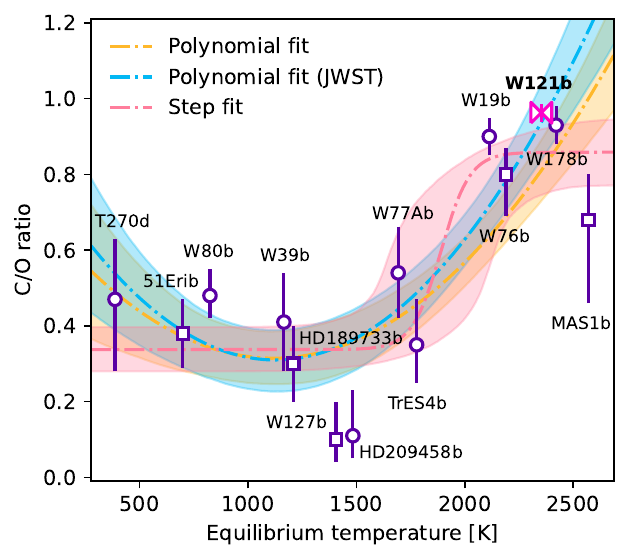}
\caption{Known and well-constrained C/O ratios for a sample of exoplanets spanning a wide range of equilibrium temperatures are shown, alongside the precise C/O ratio of WASP-121b derived in this work (highlighted as a magenta bowtie). Values based on JWST observations are marked as circles, while those from other facilities are shown as squares. The plot reveals a clear trend of super-solar C/O ratios among the UHJ population. To investigate potential correlations between C/O ratios and equilibrium temperatures, model fits using a quadratic polynomial (applied to the full dataset and a JWST-only subset), and a smooth step function are shown, each with shaded 1$\sigma$ confidence intervals. \label{fig:fig4}}
\end{figure}

To investigate a potential correlation between the C/O ratio and planetary equilibrium temperature, we compiled a sample of well-constrained C/O measurements spanning a broad range of equilibrium temperatures, as shown in Figure \ref{fig:fig4}. Literature values were taken from \citep{2024arXiv240303325B, 2023A&A...673A..98B, 2024AJ....168...14W, 2024ApJ...963L...5X, 2024ApJ...962L..30E, 2024AJ....167...43F, 2023MNRAS.525.2985R, 2023MNRAS.522.5062B, 2011Natur.469...64M, 2024A&A...685A..64K, 2025arXiv250601800W, 2025MNRAS.539.1381M, saha2025a, saha2025b}, with some equilibrium temperatures adopted from \citet{2023ApJS..268....2S, 2024ApJS..274...13S}. We modeled the distribution of C/O ratios as a quadratic function of equilibrium temperature for both the full dataset and a JWST-only subset using the form:

\[
\mathrm{C/O} = \mathrm{a} \, \mathrm{T}_{\mathrm{eq}}^2 + \mathrm{b} \, \mathrm{T}_{\mathrm{eq}} + \mathrm{c}
\]

The best-fit coefficients for the full dataset were $\mathrm{a} = 0.72 \pm 0.27$, $\mathrm{b} = (-7.27 \pm 3.84) \times 10^{-4}$, and $\mathrm{c} = (3.25 \pm 1.23) \times 10^{-7}$, while for the JWST-only subset, we obtained $\mathrm{a} = 0.84 \pm 0.27$, $\mathrm{b} = (-9.41 \pm 4.09) \times 10^{-4}$, and $\mathrm{c} = (4.18 \pm 1.38) \times 10^{-7}$ (see Figure \ref{fig:fig4}). In both cases, the well-constrained fit parameters (with low degeneracy) support the presence of an emerging trend, with JWST-driven measurements dominating the fit due to their significantly higher precision and accuracy.

To further investigate the possible onset of physical or chemical processes responsible for the elevated C/O ratios observed in UHJs compared to cooler populations, we also fit the C/O—equilibrium temperature relation using a smooth step function of the form:

\[
\mathrm{C/O} = \mathrm{C/O}_{\mathrm{low}} + \frac{\mathrm{C/O}_{\mathrm{high}} - \mathrm{C/O}_{\mathrm{low}}}{1 + \exp\left( -\frac{T - T_{\mathrm{break}}}{k} \right)}
\]

The best-fit parameters were $T_{\mathrm{break}} = 1890.21 \pm 256.0~\mathrm{K}$, $\mathrm{C/O}_{\mathrm{low}} = 0.337 \pm 0.059$, and $\mathrm{C/O}_{\mathrm{high}} = 0.859 \pm 0.087$. The relatively small uncertainties on these parameters support the presence of a significant transition in C/O ratios near the break temperature, indicative of a sharp increase in C/O above $T_{\mathrm{break}}$. Continued efforts to expand the sample with precisely measured C/O ratios—both for UHJs and for cooler planets—through upcoming large-scale JWST programs will be crucial for constraining the nature and origin of this apparent transition.

Using the abundances estimated from the free chemistry retrieval of the \texttt{Eureka!} spectra, we derived a super-solar metallicity of 4.7$_{-1.38}^{+1.99}$ $\times$solar, indicating possible photoevaporation signatures in the ultra-hot day-side atmosphere \citep[e.g.,][]{2018ApJ...865...75N, 2020arXiv200508676M}. Alternatively, the planet may have migrated inwards during the early stages of its evolution \citep[e.g.,][]{2019A&A...627A.127C}, accreting substantial amounts of metal-rich dust from the inner protoplanetary disk \citep[e.g.,][]{1998AREPS..26...53B, 2021MNRAS.503.5254L, 2023A&A...675A.176U}. From these abundances, we also estimated Si/O to be 0.034$\pm$0.024, a sub-solar value, and Si/H to be 2.34$_{-1.18}^{2.0}$ $\times$solar. However, Ti/H and V/H constrained by TiO and VO abundances from the free chemistry retrievals remain significantly lower than the solar values, suggesting that heavy refractory species may be sequestered in these atmospheres despite their extremely high day-side temperatures.

Our findings provide a broader and much-needed window into the atmospheric chemistry and physical processes of the unique and valuable EUHJ WAP-121b, establishing it as one of the most comprehensively characterized exoplanetary atmospheres to date. These results offer crucial benchmarks for testing the next-generation models of planetary evolution and atmospheric dynamics, advancing our understanding of the physical and chemical mechanisms shaping such extreme environments. Ultimately, a population-level analysis of similarly well-characterized exoplanet atmospheres will yield vital insights into the origin and evolution of UHJs and the broader exoplanet population. \linebreak\linebreak

We sincerely thank the scientific reviewer for her constructive comments and feedback on the manuscript. SS acknowledges Fondo Comité Mixto-ESO Chile ORP 025/2022 to support this research. The computations presented in this work were performed using the Geryon-3 supercomputing cluster, which was assembled and maintained using funds provided by the ANID-BASAL Center FB210003, Center for Astrophysics and Associated Technologies, CATA. This work is based [in part] on observations made with the NASA/ESA/CSA James Webb Space Telescope. The data were obtained from the Mikulski Archive for Space Telescopes at the Space Telescope Science Institute, which is operated by the Association of Universities for Research in Astronomy, Inc., under NASA contract NAS 5-03127 for JWST. These observations are associated with program $\#$2055.  JSJ gratefully acknowledges support by FONDECYT grant 1240738 and from the ANID BASAL project FB210003.

\bibliography{ms}

\clearpage

\section*{Appendix}

\renewcommand{\thefigure}{A\arabic{figure}}
\renewcommand{\thetable}{A\arabic{table}}
\renewcommand{\theequation}{A\arabic{equation}}
\renewcommand{\thepage}{A\arabic{page}}
\setcounter{figure}{0}
\setcounter{table}{0}
\setcounter{equation}{0}
\setcounter{page}{1}

\begin{table}[!h]
    \centering
    \caption{Priors and estimated values of the parameters used in the white light curve fit (from \texttt{Eureka}).
    \label{tab:lc_w}}
    \begin{tabular}{lll}
        \hline
        \hline
        Parameter & Prior & Value \\
        \hline
        $P$ (days) & fixed & $1.27492504$ \\
        $T_{0,1}$ (MJD) & $\mathcal{U}(60244.42,60244.62)$ & $60244.52022^{+0.00058}_{-0.00060}$ \\
        $T_{0,2}$ (MJD) & $\mathcal{U}(59867.04,59867.24)$ & $59867.14282^{+0.00029}_{-0.00029}$ \\
        $a/R_\star$ & $\mathcal{U}(2.5,5.0)$ & $3.77435^{+0.0297}_{-0.0639}$ \\
        $i$ (deg) & $\mathcal{U}(85.0,90.0)$ & $88.20^{+1.24}_{-1.66}$ \\
        $R_{p,1}/R_\star$ & fixed & $0.12355$ \\
        $R_{p,2}/R_\star$ & fixed & $0.12355$ \\
        $R_{p,3}/R_\star$ & fixed & $0.12355$ \\
        $F_{p,1}/F_\star$ & $\mathcal{U}(0.0,0.01)$ & $0.001158^{+0.000047}_{-0.000051}$ \\
        $F_{p,2}/F_\star$ & $\mathcal{U}(0.0,0.01)$ & $0.003886^{+0.000115}_{-0.000129}$ \\
        $F_{p,3}/F_\star$ & $\mathcal{U}(0.0,0.01)$ & $0.004817^{+0.000147}_{-0.000173}$ \\
        $e$ & fixed & $0.0$ \\
        $\omega$ (deg) & fixed & $90.0$ \\
        \hline
        $a_{0,1}$ & $\mathcal{N}(1,0.01)$ & $0.99915^{+0.00007}_{-0.00006}$ \\
        $a_{1,1}$ & $\mathcal{N}(0,0.001)$ & $-0.00002^{+0.00011}_{-0.00012}$ \\
        $a_{2,1}$ & $\mathcal{N}(0,0.001)$ & $-0.00027^{+0.00016}_{-0.00015}$ \\
        $a_{3,1}$ & $\mathcal{N}(0,0.001)$ & $0.00005^{+0.00013}_{-0.00014}$ \\
        $a_{4,1}$ & $\mathcal{N}(0,0.001)$ & $-0.00000^{+0.00005}_{-0.00005}$ \\
        $b_{0,1}$ & $\mathcal{N}(1,0.01)$ & $0.99922^{+0.00007}_{-0.00006}$ \\
        $b_{1,1}$ & $\mathcal{N}(0,0.001)$ & $-0.00007^{+0.00011}_{-0.00010}$ \\
        $b_{2,1}$ & $\mathcal{N}(0,0.001)$ & $-0.00021^{+0.00016}_{-0.00014}$ \\
        $b_{3,1}$ & $\mathcal{N}(0,0.001)$ & $-0.00009^{+0.00014}_{-0.00013}$ \\
        $b_{4,1}$ & $\mathcal{N}(0,0.001)$ & $-0.00004^{+0.00006}_{-0.00005}$ \\
        \hline
        $a_{0,2}$ & $\mathcal{N}(1,0.01)$ & $0.99704^{+0.00011}_{-0.00010}$ \\
        $a_{1,2}$ & $\mathcal{N}(0,0.001)$ & $-0.00030^{+0.00013}_{-0.00013}$ \\
        $a_{2,2}$ & $\mathcal{N}(0,0.001)$ & $-0.00029^{+0.00020}_{-0.00018}$ \\
        $a_{3,2}$ & $\mathcal{N}(0,0.001)$ & $0.00004^{+0.00020}_{-0.00020}$ \\
        $a_{4,2}$ & $\mathcal{N}(0,0.001)$ & $-0.00004^{+0.00014}_{-0.00014}$ \\
        $b_{0,2}$ & $\mathcal{N}(1,0.01)$ & $0.99617^{+0.00012}_{-0.00011}$ \\
        $b_{1,2}$ & $\mathcal{N}(0,0.001)$ & $-0.00011^{+0.00017}_{-0.00015}$ \\
        $b_{2,2}$ & $\mathcal{N}(0,0.001)$ & $-0.00038^{+0.00024}_{-0.00021}$ \\
        $b_{3,2}$ & $\mathcal{N}(0,0.001)$ & $0.00010^{+0.00023}_{-0.00022}$ \\
        $b_{4,2}$ & $\mathcal{N}(0,0.001)$ & $0.00003^{+0.00012}_{-0.00012}$ \\
        \hline
        $a_{0,3}$ & $\mathcal{N}(1,0.01)$ & $0.99551^{+0.00014}_{-0.00012}$ \\
        $a_{1,3}$ & $\mathcal{N}(0,0.001)$ & $-0.00013^{+0.00016}_{-0.00017}$ \\
        $a_{2,3}$ & $\mathcal{N}(0,0.001)$ & $-0.00035^{+0.00026}_{-0.00022}$ \\
        $a_{3,3}$ & $\mathcal{N}(0,0.001)$ & $0.00003^{+0.00024}_{-0.00026}$ \\
        $a_{4,3}$ & $\mathcal{N}(0,0.001)$ & $-0.00002^{+0.00017}_{-0.00018}$ \\
        $b_{0,3}$ & $\mathcal{N}(1,0.01)$ & $0.99592^{+0.00016}_{-0.00014}$ \\
        $b_{1,3}$ & $\mathcal{N}(0,0.001)$ & $0.00002^{+0.00021}_{-0.00019}$ \\
        $b_{2,3}$ & $\mathcal{N}(0,0.001)$ & $-0.00041^{+0.00031}_{-0.00027}$ \\
        $b_{3,3}$ & $\mathcal{N}(0,0.001)$ & $0.00010^{+0.00029}_{-0.00028}$ \\
        $b_{4,3}$ & $\mathcal{N}(0,0.001)$ & $0.00004^{+0.00016}_{-0.00016}$ \\
        \hline
        \hline
    \end{tabular}
\end{table}

\begin{figure}
\centering
\includegraphics[width=0.5\columnwidth]{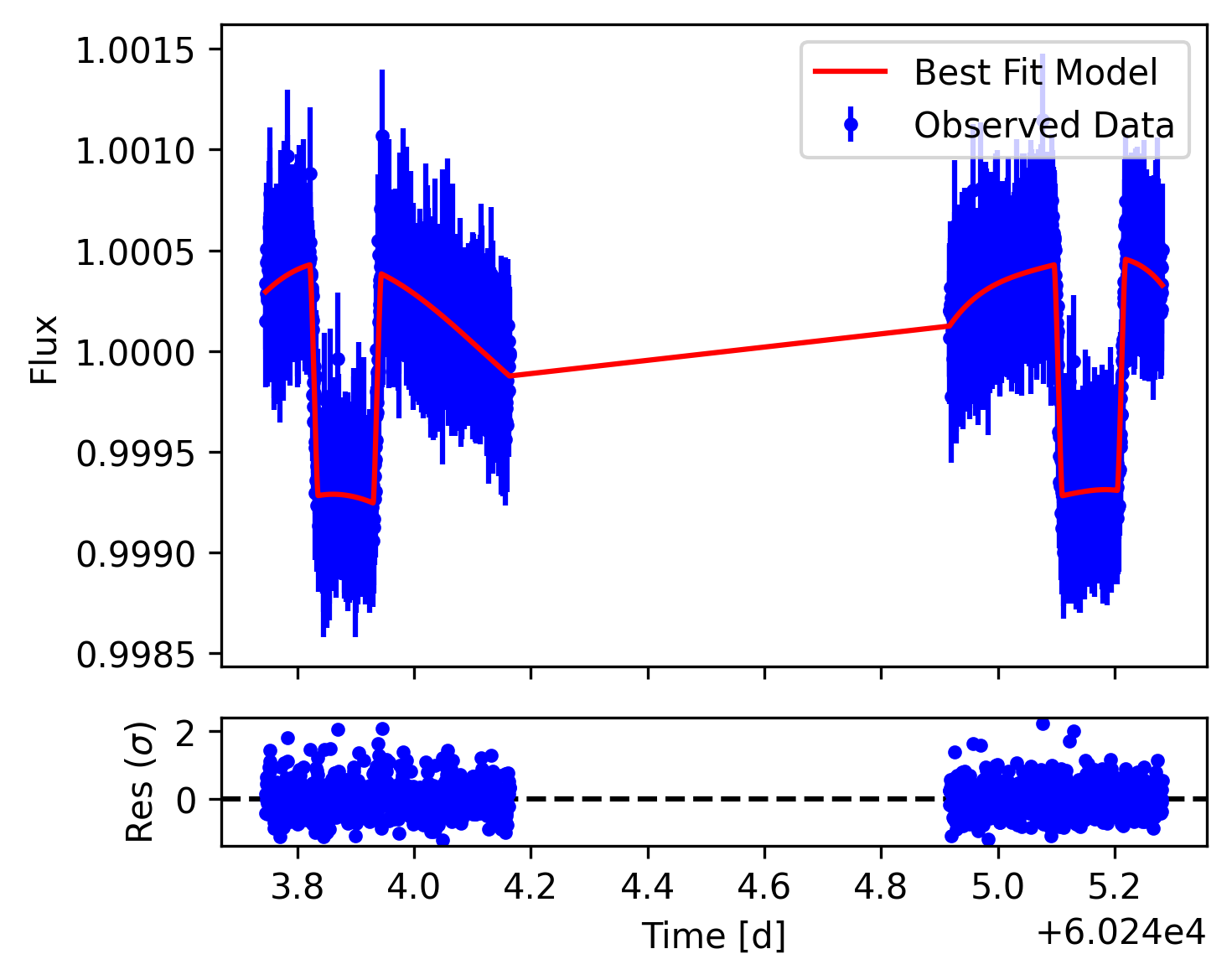}
\includegraphics[width=0.5\columnwidth]{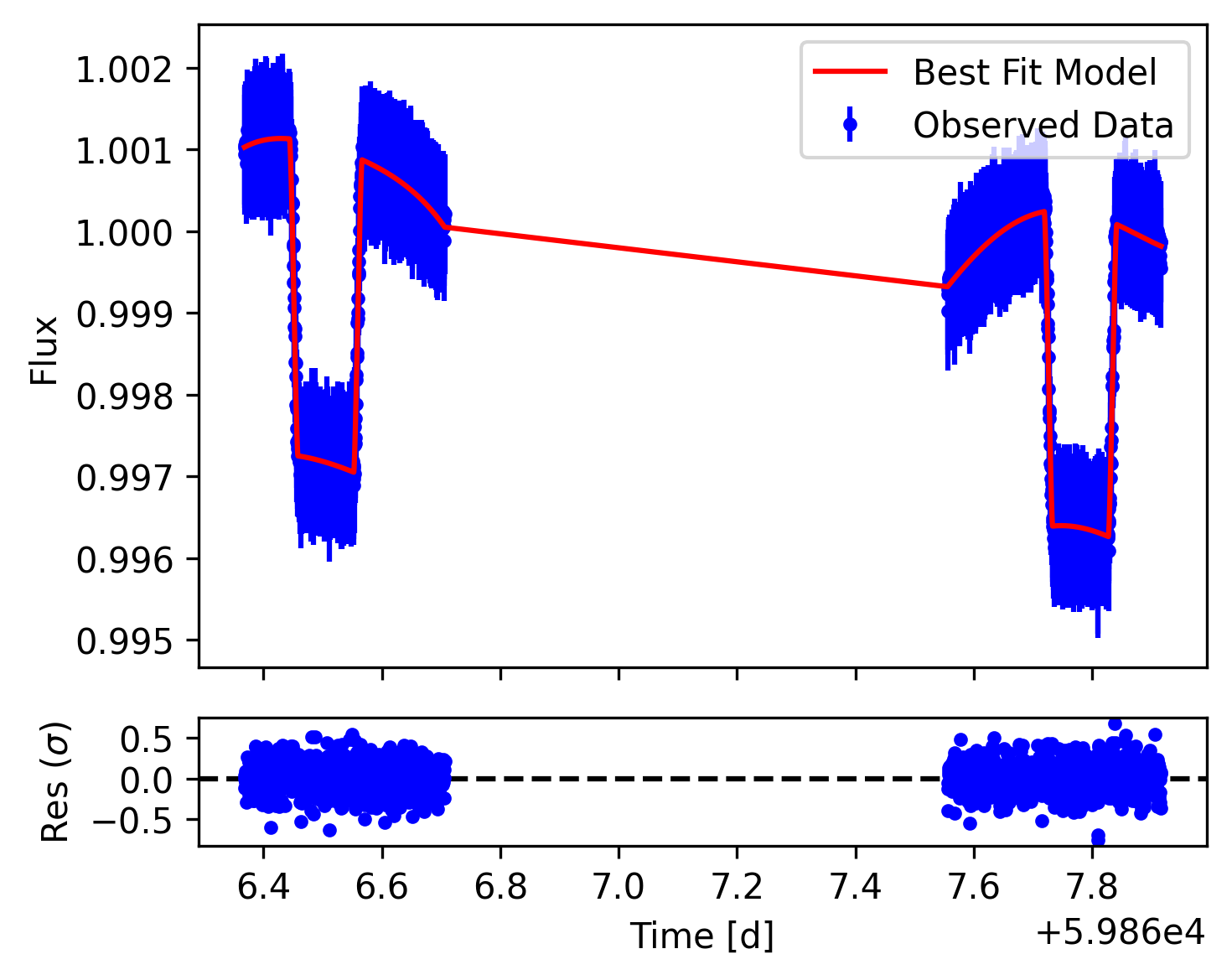}
\includegraphics[width=0.5\columnwidth]{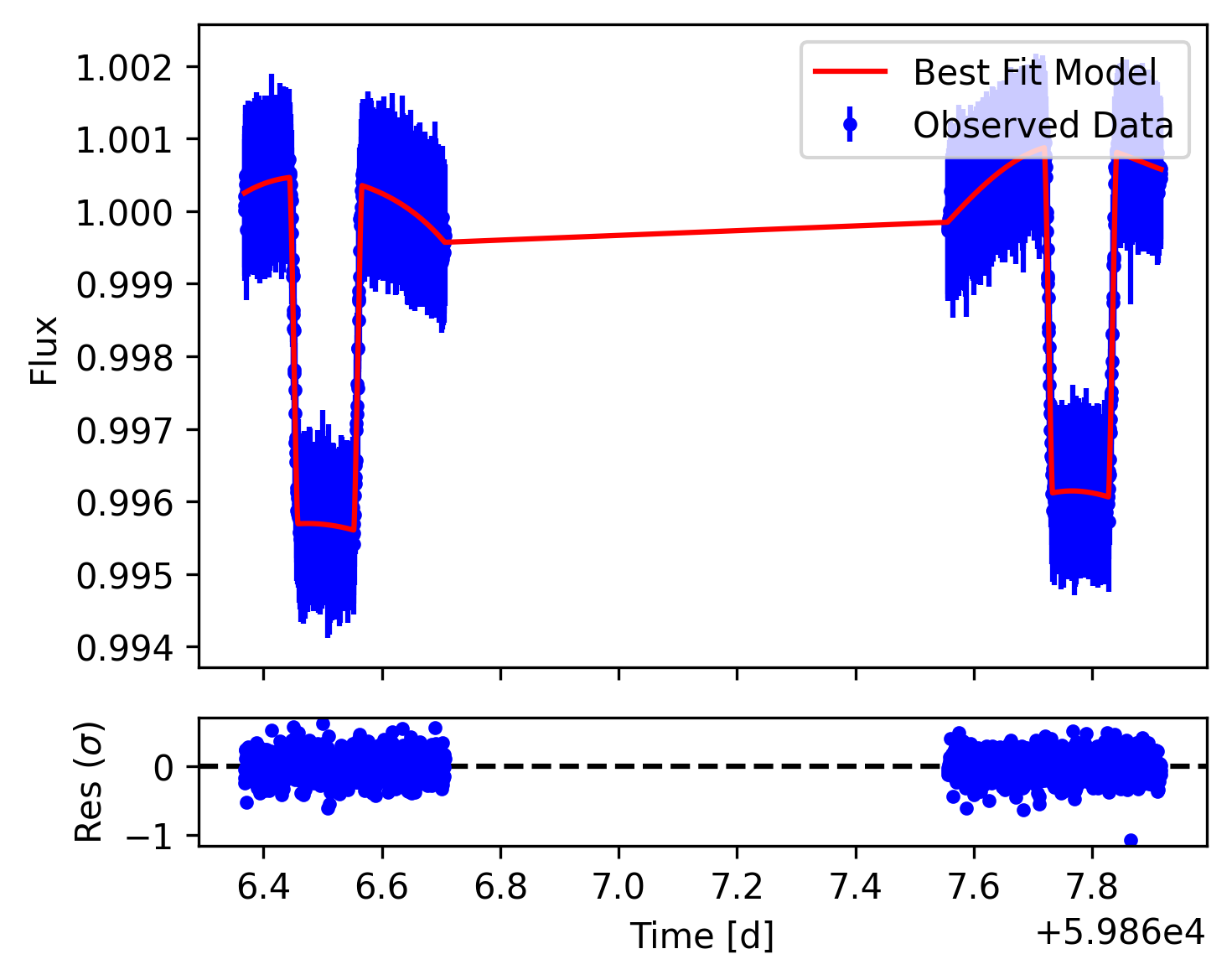}
\caption{The white light curves from NIRISS (Order 1), and NIRSpec/G395H NRS1 and NRS2, shown from top to bottom, respectively, along with the best-fit secondary eclipse models including polynomial detrending. The corresponding residuals are displayed beneath each light curve. \label{fig:lc_w}}
\end{figure}

\begin{figure}
\centering
\includegraphics[width=0.5\columnwidth]{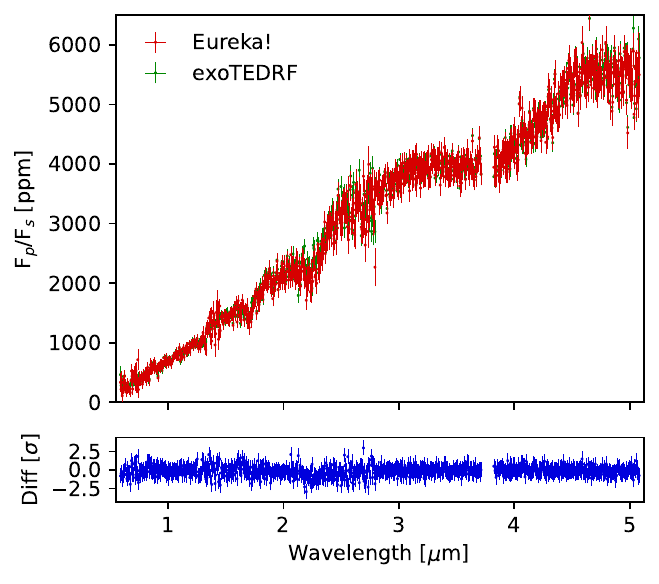}
\caption{Emission spectra obtained from both \texttt{Eureka!} and \texttt{exoTEDRF} are overplotted for comparison, along with their relative difference. The mean absolute difference between the two spectra is $\sim$0.5$\sigma$, indicating excellent statistical agreement. \label{fig:spec_diff}}
\end{figure}

\begin{table*}
    \centering
    \caption{Prior distributions for the free parameters used in the free chemistry atmospheric retrievals. \label{tab:retrieval_priors}}
    \begin{tabular}{l c}
        \hline
        \hline
        Parameter & Prior \\
        \hline
        $T_{\mathrm{equ}}$ & $\mathcal{U}(1500, 3500)$ \\
        $T_{\mathrm{int}}$ & $\mathcal{U}(0, 1000)$ \\
        $\gamma$ & $\mathcal{U}(0.01, 5.0)$ \\
        $\log \kappa_{\mathrm{IR}}$ & $\mathcal{U}(-4, 0)$ \\
        $f_{\mathrm{sed}}$ & $\mathcal{U}(0.1, 10)$ \\
        $\sigma$ & $\mathcal{U}(1.5, 3)$ \\
        $\log K_{zz}$ & $\mathcal{U}(2, 15)$ \\
        $f_c$ (cloud fraction) & $\mathcal{U}(0, 1)$ \\
        $[\mathrm{H_2O}]$ & $\mathcal{U}(-12, -1)$ \\
        $[\mathrm{CO}]$ & $\mathcal{U}(-12, -0.5)$ \\
        $[\mathrm{SiO}]$ & $\mathcal{U}(-12, -0.5)$ \\
        $[\mathrm{HCN}]$ & $\mathcal{U}(-14, -1)$ \\
        $[\mathrm{TiO}]$ & $\mathcal{U}(-14, -3)$ \\
        $[\mathrm{VO}]$ & $\mathcal{U}(-14, -4)$ \\
        $[\mathrm{Fe(s)}]$ & $\mathcal{U}(-14, -1)$ \\
        $[\mathrm{MgSiO_3(s)}]$ & $\mathcal{U}(-14, -1)$ \\
        $[\mathrm{Al_2O_3(s)}]$ & $\mathcal{U}(-14, -1)$ \\
        $[\mathrm{SiO_2(s)}]$ & $\mathcal{U}(-14, -1)$ \\
        $[\mathrm{TiO_2(s)}]$ & $\mathcal{U}(-14, -1)$ \\
        $[\mathrm{CaTiO_3(s)}]$ & $\mathcal{U}(-14, 0)$ \\
        \hline
        \hline
    \end{tabular}
\end{table*}

\begin{figure}
\centering
\includegraphics[width=0.5\columnwidth]{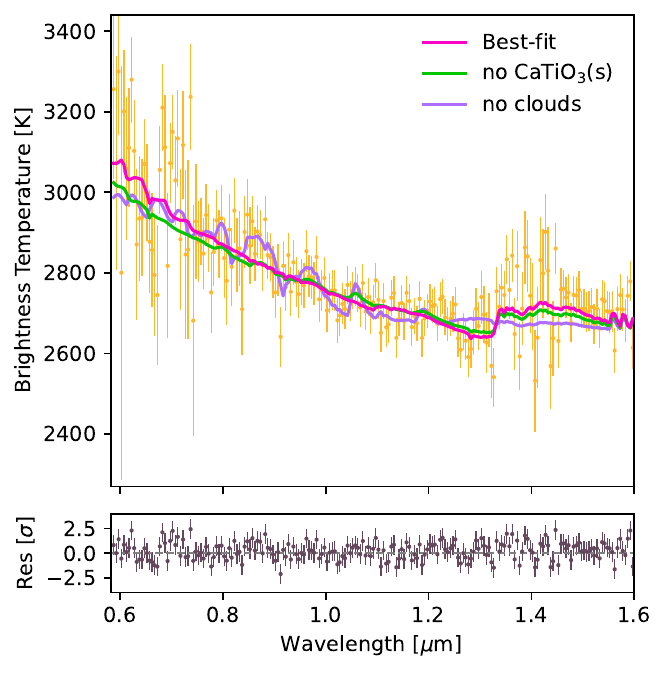}
\caption{A zoomed in version of Figure \ref{fig:fig2} at shorter wavelengths. \label{fig:fig2a}}
\end{figure}

\begin{figure}
\centering
\includegraphics[width=\columnwidth]{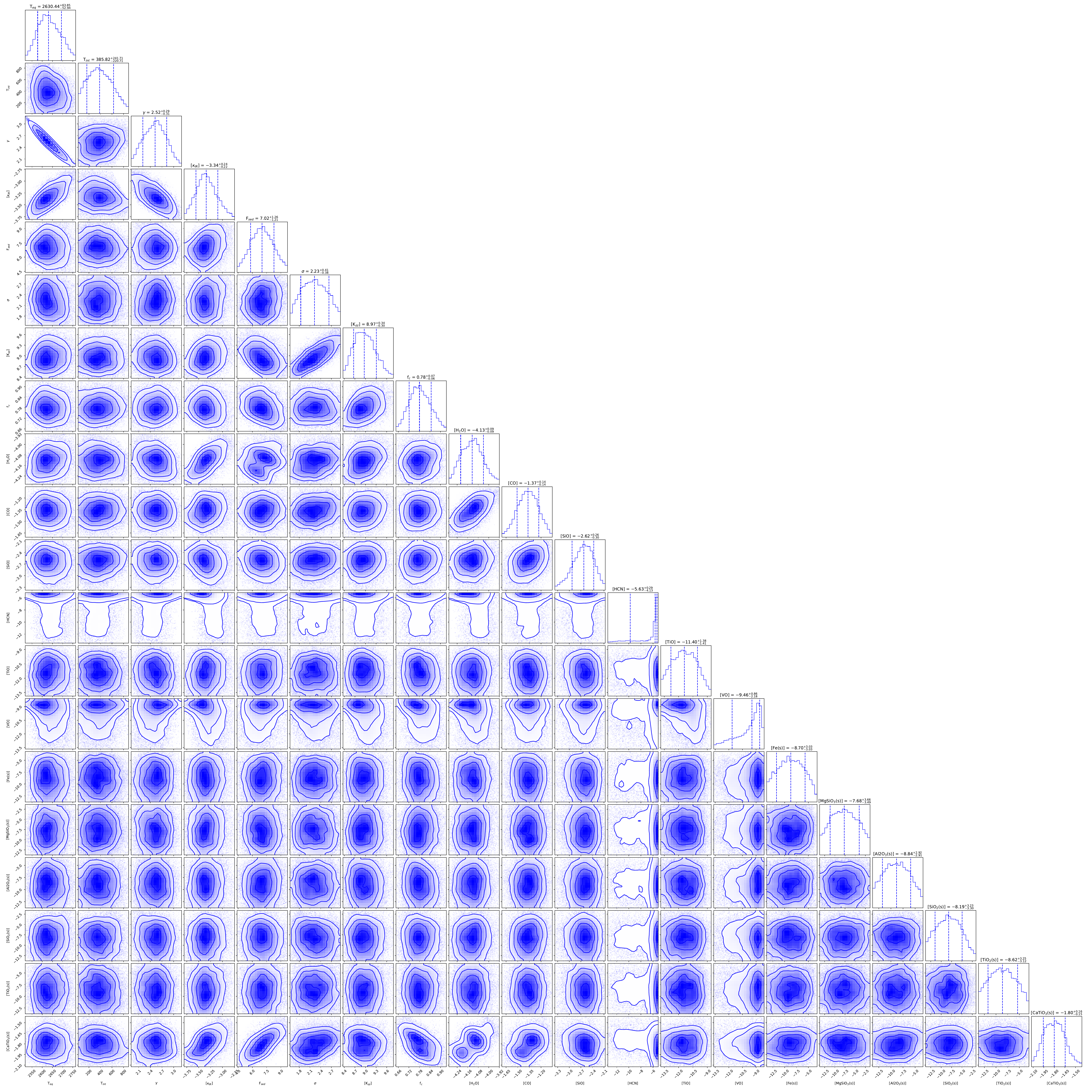}
\caption{Posterior distribution of the model parameters from the free chemistry retrieval of the \texttt{Eureka!} spectra. \label{fig:cor_eu_f}}
\end{figure}

\begin{figure}
\centering
\includegraphics[width=\columnwidth]{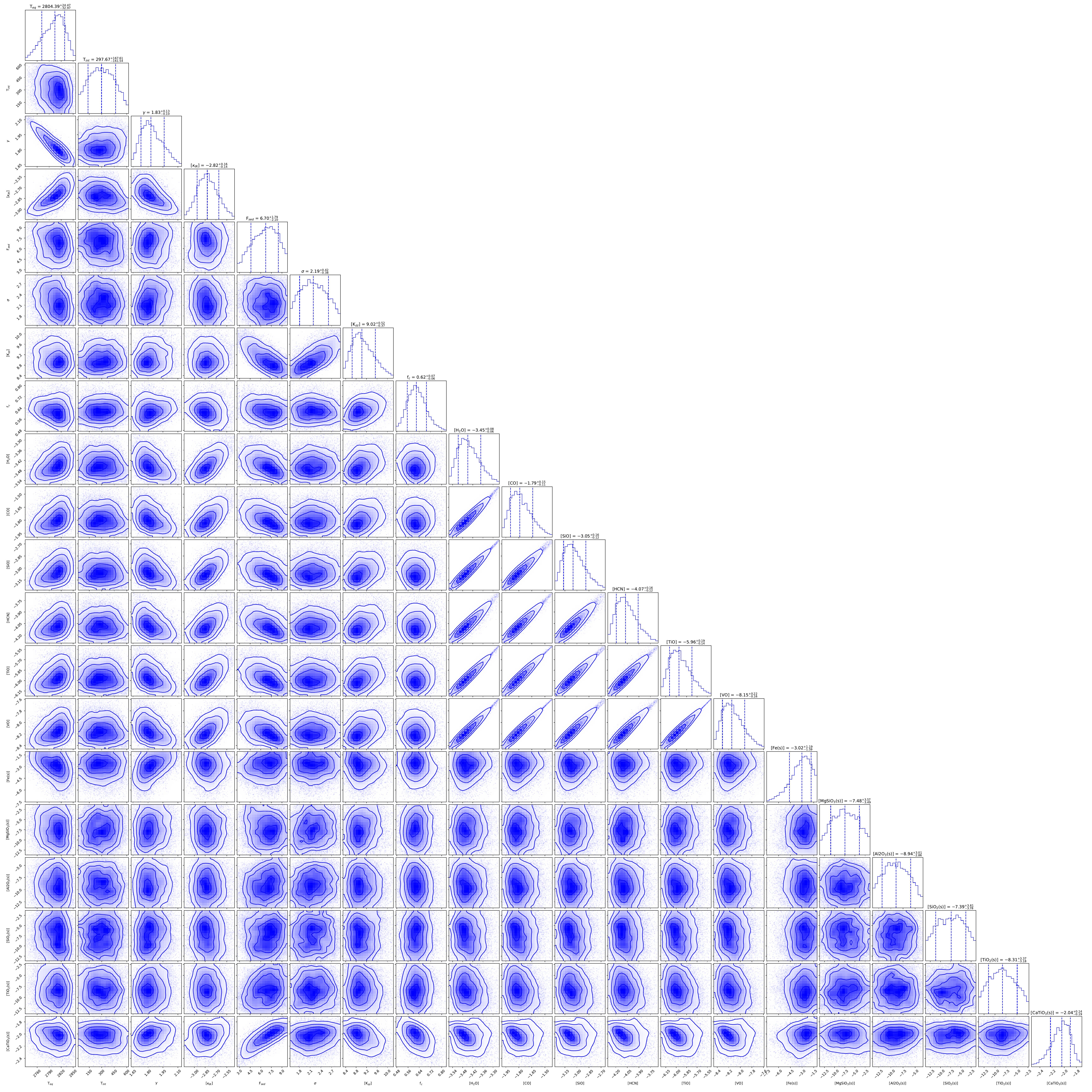}
\caption{Posterior distribution of the model parameters from the equilibrium chemistry retrieval of the \texttt{Eureka!} spectra. \label{fig:cor_eu_eq}}
\end{figure}

\begin{figure}
\centering
\includegraphics[width=\columnwidth]{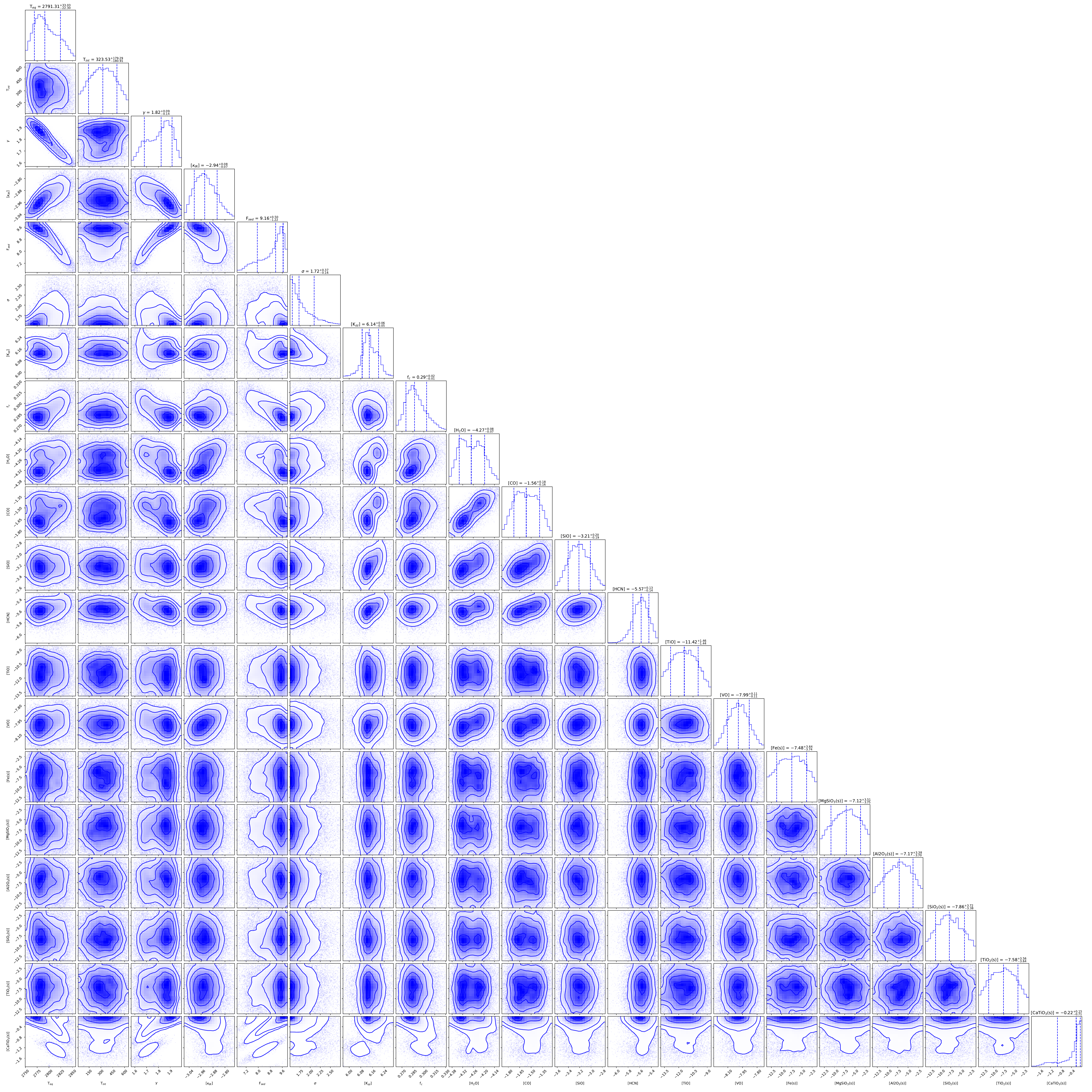}
\caption{Posterior distribution of the model parameters from the free chemistry retrieval of the \texttt{exoTEDRF} spectra. \label{fig:cor_et_f}}
\end{figure}

\begin{figure}
\centering
\includegraphics[width=\columnwidth]{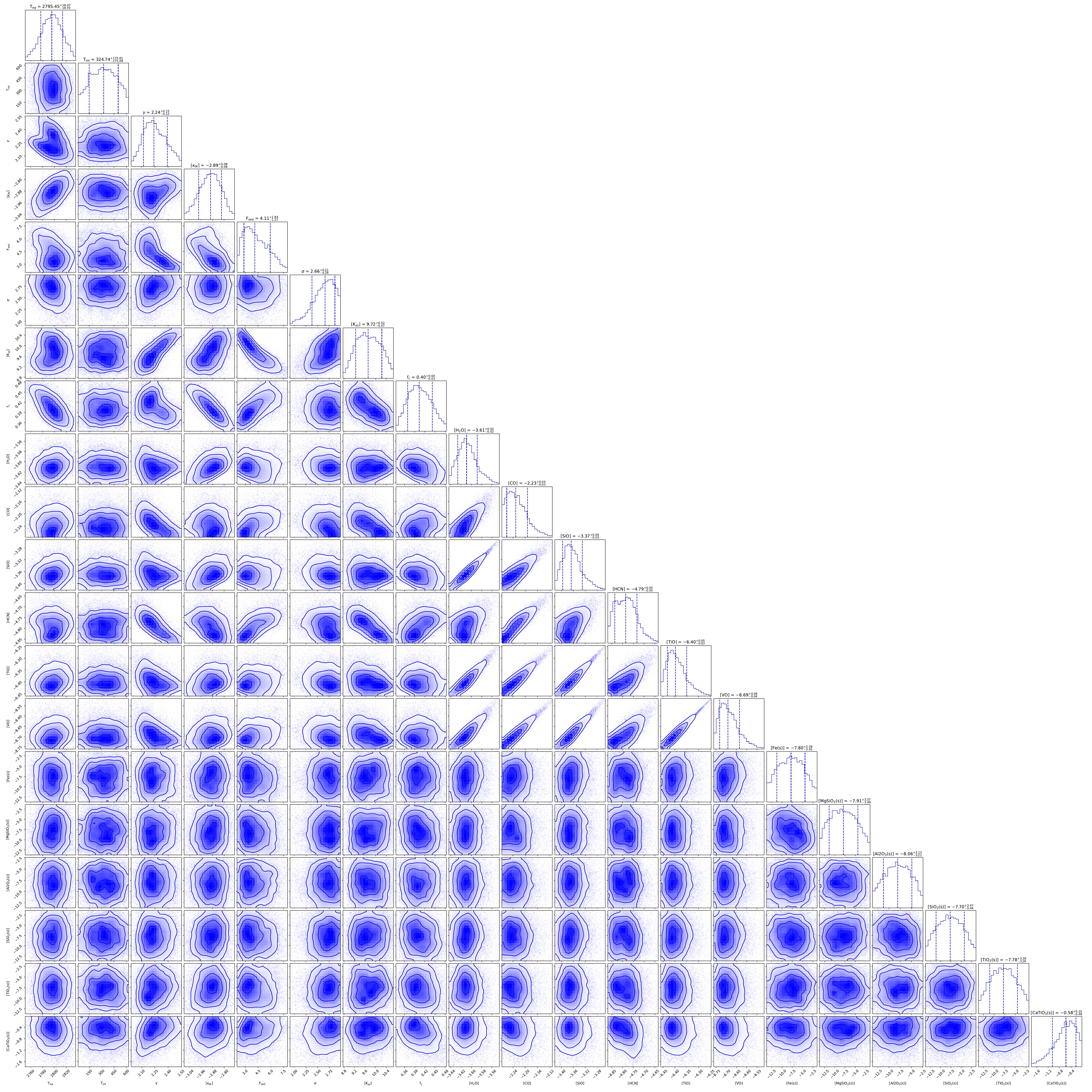}
\caption{Posterior distribution of the model parameters from the equilibrium chemistry retrieval of the \texttt{exoTEDRF} spectra. \label{fig:cor_et_eq}}
\end{figure}

\end{document}